%% file: asme2ej.tex
\documentclass[twocolumn,10pt]{wmrDoc}
\usepackage{tabularx}
\usepackage{epsfig}
\usepackage{graphicx}
\usepackage{amsmath,amssymb}
\usepackage{booktabs}
\usepackage{multirow}
\usepackage{array}
\usepackage{siunitx}
\usepackage{hyperref}
\usepackage[font=small,labelfont=bf]{caption}
\usepackage{placeins}
\hypersetup{colorlinks=true,linkcolor=black,citecolor=black,urlcolor=black}

\usepackage{pgfplots}
\usepackage{float}
\pgfplotsset{compat=1.16}

\newcommand{\EAUuOWD}{3.51}
\newcommand{\EACoreInetOWDedge}{5.125}     
\newcommand{\EACoreInetOWDcentral}{17.91}  
\newcommand{\EAUPFproc}{0.035}
\newcommand{\EAPCSradio}{5.0}

\newcommand{\EADNSwide}{45.0}      
\newcommand{\EADNSedge}{1.9}       
\newcommand{\EARTTedge}{17.27}     
\newcommand{\EATLSfull}{17.27}     
\newcommand{\EAPCSsetupBest}{80.0} 
\newcommand{\EAPCSdiscA}{20.0}     
\newcommand{\EAPCSdiscB}{40.0}     

\newcommand{\WPuUuOneWay}{\SI{3.51}{\milli\second}}          

\newcommand{\WPDNSMobileMedian}{\SIrange{35}{60}{\milli\second}} 

\title{Latency Components in 5G for Edge Application Discovery and Proximity Services:\\
Targets, Measurements, and Practical Working Points}

\author{David Rico
    \affiliation{
        drico@pa.uc3m.es
    }
}

\begin{document}
\maketitle

\begin{abstract}
This technical note surveys latency contributors that matter when deploying edge-enabled applications and proximity services in 5G.
Rather than proposing a new mechanism, we focus on building a reproducible latency catalogue grounded on: (i) theoretical targets and ranges
reported in standards, and (ii) representative empirical measurements from the literature (or explicit derivations from such measurements).
We cover user-plane components (radio access, transport and core traversal, UPF processing, and direct sidelink data paths) as well as discovery and setup
contributors (DNS and EASDF behavior, inter-NF signaling costs in the service-based core, proximity discovery timing, and security/link establishment).
For several control-plane functions, standards define procedures and APIs but do not fix execution time in milliseconds; in these cases we document the gap
and report only ranges or proxy instantiations with clear assumptions.
\end{abstract}

\section{Introduction}
The transition from centralized cloud connectivity toward a compute continuum has made latency budgeting a first-order concern for edge applications.
Even when compute is placed close to the radio access, end-to-end delays depend on multiple subsystems: radio scheduling and processing, transport and core
traversal, edge traffic steering, name resolution, and (when relevant) proximity discovery and link setup.
A recurring problem in infrastructure-only baselines is the circuitous path created by user-plane anchoring and transport traversal (often referred to as
tromboning or hair-pinning), which can dominate delays even when endpoints are physically close.

This note consolidates a set of latency values and ranges that are commonly needed for back-of-the-envelope analysis.
We intentionally separate: (i) values that are normatively stated as targets/ranges in standards, (ii) values measured in real deployments or testbeds,
and (iii) values that, at the best of our knowledge, can only be obtained by deriving from measured breakdowns or by selecting a working point within a documented range.

\section{Scope and methodology}
\label{sec:scope}

This survey targets \emph{5G systems in general} (FR1/sub-6 and FR2/mmWave, NSA and SA),
and focuses on latency components that matter for:
(i) Edge Application Server (EAS) discovery in 5G, and
(ii) Proximity Services (ProSe) discovery and data exchange over PC5.

Although part of the empirical anchor used in this survey comes from a 5G mmWave measurement study,
we do \emph{not} use it to claim that ``all 5G'' has the same radio latency.
Instead, we use mmWave as a representative \emph{edge-proximal deployment case} that provides
a verified decomposition of RTT into (a) 5G RAN and (b) 5G Core+Internet contributions.
Those latter contributions primarily depend on deployment topology and server placement rather than on carrier
frequency.

To avoid biasing the survey towards FR2 only, we additionally include operational measurements from
sub-6 5G deployments, which show that end-to-end latency commonly lies in the tens-of-ms regime when
the destination is not edge-proximal.

\section{Taxonomy}
\subsection{What counts as ``theoretical'' vs ``empirical''}
We use ``theoretical'' to denote standard-driven targets/ranges (e.g., latency objectives in 3GPP studies).
We use ``empirical'' to denote values measured in experiments, field trials, or detailed evaluations (e.g., RTT breakdown measurements, EASDF latency under load, or UPF forwarding microbenchmarks).
Whenever only RTT is available, we state the conversion assumptions used to infer one-way components.

\subsection{Component taxonomy}
We structure contributors into two groups.

\textbf{User-plane components}:
\begin{itemize}
  \item Uu radio access latency (uplink and downlink).
  \item Transport and core traversal for infrastructure routing (including hair-pinning sensitivity).
  \item UPF processing delay.
  \item Direct sidelink/PC5 data-path latency.
\end{itemize}

\textbf{Discovery and setup components}:
\begin{itemize}
  \item DNS resolution latency (wide-area baselines and edge-local/EASDF cases).
  \item Inter-NF signaling in the service-based core (SBA/SBI) as a contributor to control-plane procedures.
  \item Proximity discovery timing (announcement-based vs solicitation-based behaviors).
  \item Security and link establishment overhead.
\end{itemize}

\subsection{One-way conversion conventions}
Some sources report RTT decompositions into ``RAN'' and ``Core+Internet'' contributions.
For a practical one-way decomposition:
\begin{itemize}
  \item We approximate Uu one-way latency as RTT$_{\text{RAN}}/2$.
  \item We treat the ``Core+Internet'' term as a practical proxy for transport and core traversal in infrastructure baselines; this is deployment-dependent,
        but it captures the strong sensitivity to server placement observed in field measurements.
\end{itemize}
These conventions follow the explicit reasoning documented in the mmWave breakdown analysis \cite{mmwave_pam23}.

\section{User-plane latency components}
\subsection{Uu radio access: targets and measured breakdowns}
\subsubsection{Standard-driven targets}
3GPP TR 38.913 reports widely used user-plane latency objectives for different service families, such as URLLC and eMBB.
These values are typically interpreted as design targets under idealized assumptions \cite{3gpp_tr_38913}.

\subsubsection{Empirical RTT breakdown and a practical one-way estimate}
A commercial 5G mmWave measurement study reports RTT breakdown values separating a RAN component from ``Core+Internet'' \cite{mmwave_pam23}.
For an edge placement case (WL), it reports a representative RAN RTT around \SI{7.02}{\milli\second}.
A common practical instantiation for one-way access is therefore:
\[
L_{Uu}^{UL} \approx L_{Uu}^{DL} \approx \frac{\SI{7.02}{\milli\second}}{2} = \WPuUuOneWay .
\]
This is not a standard target, but a derived estimate tied to an empirical RTT breakdown.

\subsection{Transport and core traversal: two evidence families}
\subsubsection{Model-based transport-only latencies}
A TVT model provides explicit values for transport network latency under different deployment assumptions (e.g., MEC-at-gNB versus centralized) \cite{tvt_uwicore}.
These values are useful when the analyst wants a transport-only term separated from broader ``Core+Internet'' effects.

\subsubsection{Measured ``Core+Internet'' sensitivity to placement}
The mmWave breakdown study reports a ``Core+Internet'' RTT contribution that grows strongly with server placement \cite{mmwave_pam23}.
Representative reported means include approximately:
\begin{itemize}
  \item WL (edge-like placement): \SI{10.25}{\milli\second},
  \item LZ (local-zone): \SI{30.81}{\milli\second},
  \item RG (regional): \SI{35.82}{\milli\second}.
\end{itemize}
When instantiating a hair-pinning-sensitive baseline, these values can be used as a practical proxy for transport and core traversal costs
that appear outside the radio access component.

\subsection{UPF processing delay}
UPF processing time is not standardized in milliseconds; it depends on implementation choices (software stack, acceleration, NIC offload).
A representative industrial benchmark reports consistent microsecond-level latencies for prioritized traffic (about \SIrange{31}{38}{\micro\second}) \cite{intel_upf_priority}.
This corresponds to \SIrange{0.031}{0.038}{\milli\second} and supports the common modeling choice that UPF processing is negligible compared to
transport/core traversal in non-trivial placements.

\subsection{Direct sidelink (PC5) user-plane latency}
\subsubsection{Standard-driven ranges}
TR 38.913 includes a standard-backed range often cited around \SIrange{3}{10}{\milli\second} for direct sidelink communication profiles \cite{3gpp_tr_38913}.

\subsubsection{Experimental values depend on technology and stack}
Experimental studies on PC5-like links (e.g., C-V2X) can report higher latencies (often in the order of tens of milliseconds) depending on stack and mode.
For instance, link-level measurements have reported average delays on the order of \SI{25}{\milli\second} for C-V2X versus around \SI{10}{\milli\second} for ITS-G5 in specific configurations \cite{dlr_cv2x_measurements}.
This highlights why it is important to label whether a PC5 value is a standard target or an observed stack-dependent figure.

\section{Discovery and setup latency components}
\subsection{DNS and edge-local name resolution via EASDF}
\subsubsection{Wide-area mobile DNS baselines}
Mobile DNS lookup latency can be substantial when relying on recursive resolution across the Internet.
Measurements in LTE report median DNS lookup times typically around \WPDNSMobileMedian, varying across operators and conditions \cite{pam2019_dns}.

\subsubsection{Edge-local behavior and EASDF}
EASDF is standardized as an edge-related network function for steering and discovery support (architecturally defined, not as a latency guarantee) \cite{3gpp_ts_23548}.
The EASDF-focused evaluation in \cite{easdf_6gbricks} reports UE-experienced DNS-related latency in an edge-local setup of about \SIrange{1.8}{2.2}{\milli\second}
at low to moderate load, increasing toward higher values under very high request rates. It also reports an EASDF-added latency around \SIrange{0.85}{0.95}{\milli\second}
at low/moderate load, increasing toward a few milliseconds at the highest loads \cite{easdf_6gbricks}.
These measurements justify modeling edge-local DNS resolution in the low single-digit millisecond regime when EASDF and DNS are co-located.

\subsection{Control-plane inter-NF signaling in SBA/SBI}
Many edge-enablement and proximity procedures trigger exchanges between core NFs over the service-based interface.
A low-latency 5G core study reports that kernel-based SBI multi-hop completion times can fall in the \SIrange{5}{15}{\milli\second} range,
while optimized approaches can reduce this \cite{l25gc_sbi_latency}.
These results are relevant whenever one needs to instantiate inter-NF signaling overhead in analytical models.

\subsection{Application-layer security handshake (TLS 1.3)}
Transport and core procedures do not necessarily account for application-layer security establishment.
TLS~1.3 typically completes a full handshake in 1 RTT, while session resumption may enable 0-RTT data in warm-start conditions.
Therefore, when budgeting cold-start application setup time, it is reasonable to add a 1-RTT term on top of name resolution and basic reachability.
We explicitly separate this modelled term from measured network RTT components and from link-layer establishment times.

\subsection{PC5 link setup and security: validated order-of-magnitude}
\label{sec:pc5-setup}

3GPP specifications define the signaling procedures for PC5 direct communication (Stage 2/3),
but do not mandate universal timer values in milliseconds; the resulting setup time depends on
sidelink configuration and timer tuning.

A NIST simulation-based evaluation (LTE ProSe UE-to-network relay context) reports that the
sidelink period can be configured between 40~ms and 320~ms, and that establishing a secured
relay-remote connection takes slightly over 3 sidelink periods (best case) and up to 5.3 periods
(worst case). This translates to approximately 120~ms (3$\times$40~ms) up to about 1.7~s (5.3$\times$320~ms)
for the considered configurations, and should be used as an order-of-magnitude reference rather
than a universal constant.\\
We also report a theoretical lower bound of \SI{80}{\milli\second} obtained as two message exchanges under the minimum sidelink period (\SI{40}{\milli\second}).
This bound is not a measured outcome of the cited study and is only meant to contextualize best-case tuning limits.

\begin{table}[H]
\caption{EdgeApp-aligned instantiation}
\label{tab:edgeapp_inst}
\centering
\scriptsize
\setlength{\tabcolsep}{4pt}
\begin{tabular}{@{}p{2.1cm}p{2.2cm}p{1.1cm}@{}}
\toprule
\textbf{Block} & \textbf{Meaning} & \textbf{Value} \\
\midrule
Uu (one-way) & Derived from RAN RTT/2 (WL) & \EAUuOWD ms \\
Core+Internet (OWD) & Edge proxy (WL), RTT/2 & \EACoreInetOWDedge ms \\
Core+Internet (OWD) & Central proxy (RG), RTT/2 & \EACoreInetOWDcentral ms \\
UPF proc & Fast-path illustrative working point & \EAUPFproc ms \\
PC5 radio & Standard-driven working point & \EAPCSradio ms \\
\midrule
DNS (SoA w/o EASDF) & Wide-area baseline & \EADNSwide ms \\
DNS (SoA +EASDF) & Edge-local (UE-experienced) & \EADNSedge ms \\
RTT to edge & 1 exchange proxy (WL) & \EARTTedge ms \\
TLS 1.3 handshake & Model: 1 RTT (cold-start) & \EATLSfull ms \\
PC5 discovery A/B & Best-case chosen operating points & \EAPCSdiscA / \EAPCSdiscB ms \\
PC5 setup & Lower bound: 2 exchanges @ 40 ms & \EAPCSsetupBest ms \\
\bottomrule
\end{tabular}
\end{table}

\section{Summary of values: targets, measurements, and a worked instantiation}
This section summarizes all concrete values that appeared in the surveyed sources and provides a single worked instantiation (a practical working point)
to help readers reproduce an example latency budget. The working point is not normative.

\subsection{Standards-driven targets and ranges}

\begin{table}[H]
\caption{Examples of standards-driven targets/ranges commonly used in latency budgeting.}
\label{tab:targets}
\centering
\small
\begin{tabular}{@{}p{2.7cm} p{3.5cm} p{1.6cm}@{}}
\toprule
\textbf{Component} & \textbf{Target/range (standard)} & \textbf{Ref.} \\
\midrule
Uu user-plane latency (URLLC/eMBB examples) & Example targets for URLLC and eMBB user-plane latency & \cite{3gpp_tr_38913} \\
PC5 user-plane latency & \SIrange{3}{10}{\milli\second} (commonly cited range for direct sidelink profiles) & \cite{3gpp_tr_38913} \\
SL period (timing driver for solicitation-based procedures) & \SIrange{40}{320}{\milli\second} & \cite{nist_timers_931313} \\
\bottomrule
\end{tabular}
\end{table}

\subsection{Empirical measurements and directly reported values}

\begin{table*}[h]
\caption{Empirical or directly reported values in the surveyed sources.}
\label{tab:empirical}
\centering
\scriptsize
\setlength{\tabcolsep}{5pt}
\renewcommand{\arraystretch}{1.15}
\begin{tabularx}{\textwidth}{l X l}
\toprule
\textbf{Component} & \textbf{Reported empirical value(s)} & \textbf{Ref.} \\
\midrule
RAN RTT breakdown (edge placement WL) &
$T_{\text{RAN}}\approx \SI{7.02}{\milli\second}$ (RTT component) &
\cite{mmwave_pam23} \\

Core+Internet RTT breakdown &
WL: \SI{10.25}{\milli\second}, LZ: \SI{30.81}{\milli\second}, RG: \SI{35.82}{\milli\second} &
\cite{mmwave_pam23} \\

UPF processing (prioritized traffic) &
\SIrange{31}{38}{\micro\second} &
\cite{intel_upf_priority} \\

DNS lookup (mobile baseline) &
Median \WPDNSMobileMedian &
\cite{pam2019_dns} \\

EASDF edge-local DNS behavior &
UE-experienced \SIrange{1.8}{2.2}{\milli\second} (low/moderate load), EASDF-added \SIrange{0.85}{0.95}{\milli\second} (low/moderate load) &
\cite{easdf_6gbricks} \\

SBI multi-hop signaling (control plane) &
\SIrange{5}{15}{\milli\second} (kernel-based, multi-hop) &
\cite{l25gc_sbi_latency} \\

PC5 secured setup (order of magnitude) &
3--5.3 sidelink periods with SL period \SIrange{40}{320}{\milli\second} (i.e., \SI{120}{\milli\second} to \SI{1.7}{\second} in that study) &
\cite{nist_timers_931313} \\

PC5 setup lower bound (model) &
2 message exchanges $\times$ \SI{40}{\milli\second} $=$ \SI{80}{\milli\second} (theoretical best-case under minimum SL period) &
\cite{nist_timers_931313} \\

PC5 Model A example default &
$T_{\text{ann}}=\SI{1}{\second}$ (example), mean wait $\approx \SI{500}{\milli\second}$ &
\cite{ns3_prose_wns3} \\

PC5 experimental example (stack-dependent) &
C-V2X $\sim \SI{25}{\milli\second}$ vs ITS-G5 $\sim \SI{10}{\milli\second}$ (example) &
\cite{dlr_cv2x_measurements} \\
\bottomrule
\end{tabularx}
\end{table*}

\subsection{A practical working point (explicit assumptions)}
The following working point instantiates a concrete numerical example consistent with the spreadsheet guideline.
Values are either derived from empirical breakdowns (with stated rules), selected as midpoints within documented ranges, or used as proxy instantiations
when standards do not define fixed milliseconds for a function.

\begin{table*}[h]
\centering
\caption{Latency parameters used as literature-backed working points (catalogue-level, not necessarily the EdgeApp instantiation).}
\label{tab:params}
\scriptsize
\setlength{\tabcolsep}{3pt}
\renewcommand{\arraystretch}{1.15}
\resizebox{\textwidth}{!}{%
\begin{tabular}{@{}p{3.2cm}p{1.4cm}p{5.9cm}p{1.8cm}p{2.6cm}@{}}
\toprule
Component & Value used & Literature range / comment & Type & Main sources \\
\midrule
RAN RTT (WL) & 7.02 ms & mmWave edge case (WL), RTT breakdown & Measured & \cite{mmwave_pam23} \\
RAN OWD (derived) & 3.51 ms & OWD $\approx$ RTT/2 under UL/DL symmetry & Derived & \cite{mmwave_pam23} \\
Core+Internet RTT (WL) & 10.25 ms & Placement-dependent; WL/LZ/RG reported & Measured & \cite{mmwave_pam23} \\
PC5 user-plane latency & 3--10 ms & Standard-driven target range (direct sidelink profile) & Standard target & \cite{3gpp_tr_38913} \\
DNS lookup (LTE baseline) & 35--60 ms & Median DNS lookup times in mobile networks & Measured & \cite{pam2019_dns} \\
EASDF UE-experienced DNS & 1.8--2.2 ms & Edge-local setup (low/moderate load) & Measured & \cite{easdf_6gbricks} \\
SBI multi-hop signaling & 5--15 ms & Kernel-based multi-hop; optimized variants lower & Measured & \cite{l25gc_sbi_latency} \\
PC5 secured setup & 3--5.3 periods & With SL period 40--320 ms $\Rightarrow$ 120 ms--1.7 s (study range) & Measured (sim) & \cite{nist_timers_931313} \\
\bottomrule
\end{tabular}%
}
\end{table*}

\section{Gaps and limitations}
\subsection{Why some control-plane terms cannot be ``closed'' with a single ms value}
For proximity-related functions in 5GS, documents such as TS 23.304 and TS 29.555 define the function and its interfaces, but do not prescribe
a fixed execution time in milliseconds \cite{3gpp_ts_23304,3gpp_ts_29555}.
Therefore, any concrete ms value for such functions must come from a specific implementation measurement or remain a deployment-dependent model parameter.
In this note, we explicitly label these terms as proxies when we instantiate them numerically.

\subsection{On mixing targets and measurements}
Using standard targets (e.g., TR 38.913) and empirical values (e.g., mmWave breakdowns) in the same catalogue is acceptable as long as the distinction
is explicit. Targets can be interpreted as best-case design objectives, while measurements capture real-world system behavior under specific conditions.
This note intentionally reports both categories to support transparent latency budgeting.

\FloatBarrier

\input{asme2ej.bbl}

\end{document}